\documentclass[aps,pra,twocolumn,showpacs]{revtex4}
\usepackage{epsfig}
\usepackage{amsbsy,latexsym}
\usepackage{amsmath}
\usepackage{amssymb, mathrsfs}
\usepackage[mathscr]{eucal}
\usepackage{hyperref}
\usepackage{soul}

\def\<{\langle}
\def\>{\rangle}

\def\qed{$\,\blacksquare$\par}

\def\ket#1{|#1\>}
\def\bra#1{\<#1|}
\def\tr#1{{\rm tr}[#1]}

\newcommand{\hilb}[1]{\mathcal{#1}}

\newtheorem{proposition}{Proposition}

\newtheorem{theorem}{Theorem}

\def\Proof{\medskip\par\noindent{\bf Proof. }}

\newcommand{\mM}{\mathcal{M}}
\newcommand{\mN}{\mathcal{N}}

\begin{document}
\title{Optimal single shot strategies for discrimination of quantum measurements}
  \author{Michal Sedl\'ak$^{1,2}$ and M\'ario Ziman$^{2,3}$}
    \affiliation{
$^1$Department of Optics, Palack\'{y} University, 17. listopadu 1192/12, CZ-771 46 Olomouc, Czech Republic
\\ $^2$Institute of Physics, Slovak Academy of Sciences, D\'ubravsk\'a cesta 9, 845 11 Bratislava, Slovakia
\\
$^3$Faculty of Informatics,~Masaryk University,~Botanick\'a 68a,~60200 Brno,~Czech Republic}

  \date{ \today}
\begin{abstract}
We study discrimination of $m$ quantum measurements in the scenario when the unknown measurement with $n$ outcomes can be used only once. We show that ancilla-assisted discrimination procedures provide a nontrivial advantage over simple (ancilla-free) schemes for perfect distinguishability and we prove that inevitably $m \leq n$. We derive necessary and sufficient conditions of perfect distinguishability of general binary measurements. We show that the optimization of the discrimination of projective qubit measurements and their mixtures with white noise is equivalent to the discrimination of specific quantum states. In particular, the optimal protocol for discrimination of projective qubit measurements with fixed failure rate (exploiting maximally entangled test state) is described.  While minimum error discrimination of two projective 
qubit measurements can be realized without any need of entanglement, we show that discrimination of three projective qubit measurements requires a bipartite probe state. Moreover, when the measurements are not projective, the non-maximally entangled test states can outperform the maximally entangled ones.
\end{abstract}

\maketitle

\section{Introduction}
Quantum theory is statistical, hence any distinction in the performance of quantum devices is based on statistical reasoning. However, if the set of possibilities is restricted to a finite number of alternatives (communication being the best example), the observations of individual experimental outcomes represent a nontrivial information. For example, in communication Alice encodes  a letter $a$ by selecting a pre-agreed preparation procedure associated with a quantum state $\varrho_a$. In each communication round Bob is trying to estimate which preparation was selected by Alice to recover the submitted letter $a$. How often Alice and Bob succeed is the research subject of optimal state discrimination (see for instance Chapter 11 of \cite{discrreview} or \cite{reviewchefl} for an overview).

Since seminal works of Holevo and Helstrom \cite{holevo,helstrom} a lot of
research effort was invested on the various aspects of discrimination
problems, finding its applications in quantum communication, quantum
cryptography, but also in quantum computation. For instance, in its essence
the famous Grover's search algorithm \cite{grover} solves the question of optimal
and efficient discrimination of quantum oracles representing
the database elements. No doubts the discrimination problems represents
one of the central conceptual questions of quantum physics with both
practical and foundational implications. The solutions provide a natural
quantitative measures of difference, or similarity of quantum devices with
clearly justified operational meaning. Recently, the variant of discrimination
problem was used to argue the philosophical objectivity of quantum wave
function \cite{ontology}.

In comparison with the case of states \cite{intmdt0,intmdteq1,intmdt2,intmdt1}
and processes \cite{acin,dariano,sacchi1,sacchi2,wang,duan,piani,watrous,ziman1,hashimoto,chiribella}
the discrimination of quantum measurements is rather unexplored.
In Ref.~\cite{mdiscr1} authors have shown that any pair of projective
measurements can be perfectly discriminated in finite number of runs.
The question of discrimination of measurements with unlabeled outcomes
has been addressed in Refs.~\cite{ziman2, ziman3} and experimental
realizations of measurement discrimination protocols have been reported
in Refs.~\cite{obrien,mdiscr2}.

This paper addresses the question of optimal discrimination
of quantum measurements. It is organized as follows. The problem is
formulated in Sections \ref{sec:formulation} and \ref{sec:mft}. In Section
\ref{sec:perfect} we study the conditions of perfect distinguishability.
Further we continue with discrimination of quantum filters in Section
\ref{sec:qfilters}. We reduce this problem to discrimination of projective
measurements which is then investigated in Section \ref{sec:projqm}. Finally,
in Section \ref{sec:trine} we solve unambiguous discrimination of two trine
measurements demonstrating that non-maximally entangled states can outperform 
maximally entangled ones. Section \ref{sec:summary} contains the summary
of the results.

\section{Problem setting}
\label{sec:formulation}

\begin{figure}[t]
    \includegraphics[width=8cm ]{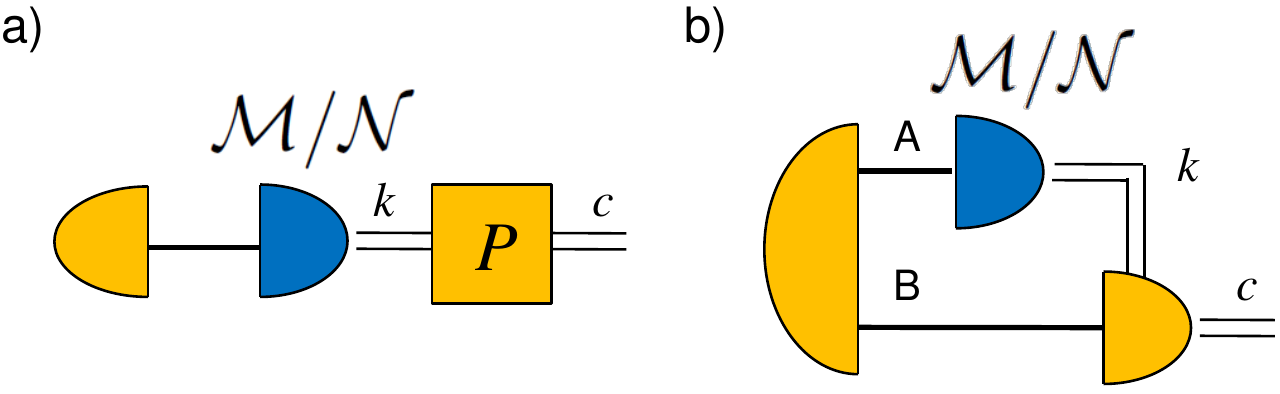}
    \caption{ \label{fig:schemes} Possible approaches to discrimination of a quantum measurement. Part $a)$ depicts the simple discrimination scheme, in which a test state $\rho$ is prepared, afterward measured with the unknown measurement device and based on the obtained outcome $k$ the measurement device is identified. Part $b)$ depicts the general discrimination scheme, in which a bipartite state is prepared; one part of it is measured with the unknown measurement device and the other part by a known ancillary measurement chosen conditionally on the actual outcome $k$ recorded in the unknown measurement. }
\end{figure}

Suppose we are given a measurement device we want to
identify, however, we can use it only once. We will assume that
a nontrivial prior knowledge on potential alternatives is given.
In the simplest case we are distinguishing among two alternatives:
the apparatus performs either a measurement $\mM$ or a measurement
$\mN$ with a priori probabilities $\eta_{\mM}$, $\eta_{\mN}$, respectively.
Our goal is to design a test that would (optimally) identify the unknown
measurement device. We will distinguish between two types of tests:
\emph{simple (ancilla-free)}
and \emph{general (ancilla-assisted)} experimental setting
(see Figure~\ref{fig:schemes}).
The simple scheme consists of the preparation of a test state
probing the unknown measurement device and of a post-processing
assignment of a conclusion for each individual measurement outcome.
In contrast, the most general (ancilla-assisted) scheme begins with
a preparation of a bipartite quantum state part of which is then
measured by the unknown measurement device. The obtained outcome $k$
is used to determine the measurement of the remaining part
of the bipartite system. Finally, based on the recorded outcomes
the guess on the identity of the unknown measurement device is made.

If the measurements $\mM$, $\mN$ are not in a specific mutual relation allowing for perfect discrimination, it is obvious that all the conclusions can not be always valid. The way how the imperfections are evaluated and processed is then used for definition of optimality. Each conclusion is characterized by the probability of being wrong (error probability). To evaluate the reliability of the conclusions of the discrimination test we use the following three quantities:
i) \emph{error probability} $p_e$ defined as the average error probability (over all conclusive outcomes);
ii) \emph{failure probability} $p_f$ given as a total
probability of inconclusive outcomes;
iii) \emph{success probability} $p_s=1-p_f-p_e$.

When all the outcomes are conclusive, i.e. $p_f=0$, and we optimize the
average success probability then we speak about \emph{minimum-error}
discrimination strategy. On the other side of the spectrum of discrimination
problems we find the \emph{unambiguous} discrimination, for which $p_e=0$, but
inconclusive outcomes are possible. The optimality is achieved when
$p_f$ is minimized. In this paper we will consider also intermediate
variations of the discrimination problems, which include
the mentioned strategies as "extremal" cases. In particular, we will consider
maximization of average success probability $p_s$ for a fixed value of
the failure probability $p_f$.

In such case we say we
implement \emph{discrimination with fixed failure rate}.
We say the measurements can be \emph{perfectly discriminated} if $p_f$
and $p_e$ can vanish simultaneously.

\section{Mathematical framework}
\label{sec:mft}

Let us denote by $\hilb{H}_d$ the $d$ dimensional Hilbert space of the quantum system under consideration. The measurement device $\mM$ is a positive operator valued measure assigning a positive operator $M_j$ for each $j\in\Omega=\{1,\ldots,n\}$ (we will not consider measurements with infinite, or uncountable number of outcomes). We can represent any measurement as a specific measure-and-prepare channel $\mM$
\begin{align}
\label{def:chsmn}
\mM(\rho)=\sum_j \tr{M_j \rho} \ket{j}\bra{j}
\end{align}
mapping states of $\hilb{H}_d$ into diagonal density operators on $\hilb{H}_n$
(probability distributions on $\Omega$), where $\hilb{H}_n$ is
$n-$dimensional Hilbert space spanned by fixed orthonormal basis $\{\ket{j}\}$.
Further, we will assume that all the measurements we want to discriminate
have the same number of outcomes.
Using this representation of measurements the problem of discrimination
can be reformulated as a special case of (quantum-classical) channel
discrimination, hence, the general results obtained for (single-shot)
discrimination of channels can be directly translated into the language
of measurements.

In what follows, we demonstrate mathematical formulation of a discrimination
problem for its simples version when the measurement device is guaranteed
to be one of two known alternatives. Generalization to
any number of measurements is straightforward. Let us denote by
$\mathcal{T}$ the test procedure we use
to discriminate between a pair of measurements $\mM$ and $\mN$
($\mN$ corresponding to POVM elements $\{N_j\}$).
We denote by $p(c|\mathcal{M},\mathcal{T})$ the conditional probability that if the measurement $\mathcal{M}$ was tested by the test procedure $\mathcal{T}$ conclusion $c \in\{\mM,\mN,f\}$ was obtained. Here $c=f$ marks that the procedure has failed and $c=\mM$, $c=\mN$ corresponds to identification of the measurement device as $\mathcal{M}$, $\mathcal{N}$, respectively. The fact that the test procedure $\mathcal{T}$ fails with fixed probability $p_f$ can be mathematically stated as:
\begin{align}
\label{def:q}
p_f= \eta_\mathcal{M}\; p(f|\mathcal{M},\mathcal{T})+\eta_\mathcal{N}\; p(f|\mathcal{N},\mathcal{T}).
\end{align}
We define
\begin{align}
p_e= \eta_\mathcal{M}\; p(\mN|\mathcal{M},\mathcal{T})+\eta_\mathcal{N}\; p(\mM|\mathcal{N},\mathcal{T}) \nonumber\\
p_s= \eta_\mathcal{M}\; p(\mM|\mathcal{M},\mathcal{T})+\eta_\mathcal{N}\; p(\mN|\mathcal{N},\mathcal{T})
\label{def:pepc}
\end{align}
the probability of error and the probability of success and clearly $p_s+p_e+p_f=1$.
Our goal is to maximize the probability of success $p_s$ for a fixed value of the failure rate $p_f$. This is equivalent to minimization of the relative error rate $p_e/(1-p_f)$ or
maximization of the relative success probability $p_s/(1-p_f)$ for fixed value
of $p_f$.

The mathematical framework for
the description of test procedures
$\mathcal{T}$ (so-called quantum testers, or process POVMs) was introduced
and developed in Refs.~\cite{zimanppovm,memeff,architecture,gutoski}.
 In this framework
the measurements are described by Choi-Jamiolkowski operators
\cite{choi,jamiolkowski}
assigned to corresponding quantum-to-classical
channels. For example, to
$\mM$ defined in Eq. (\ref{def:chsmn}) we assign
\begin{align}
\nonumber
M=(\mM\otimes\mathcal{I})[\ket{\phi_+}\bra{\phi_+}]= \sum_j \ket{j}\bra{j}\otimes M_j^T\,,
\end{align}
where $\ket{\phi_+}=\sum_k \ket{k}\otimes\ket{k}\in\hilb{H}_d\otimes\hilb{H}_d$
is the (unnormalized) maximally entangled state
($\{\ket{k}\}$ is an orthonormal basis in $\hilb{H}_d$ and
$\{\ket{j}\}$ is an orthonormal basis in $\hilb{H}_n$). Any possible
test $\mathcal{T}$ is described by a set of positive operators
$\{T_c\}$ acting on $\hilb{H}_{n}\otimes \hilb{H}_{d}$
such that
\begin{align}
\sum_c T_c =I_n\otimes \rho\,,
\label{def:norm1}
\end{align}
where $\rho$ is a density operator on
$\hilb{H}_d$, i.e.
$\rho\geq 0$ and $\tr{\rho}=1$).

Let us define projectors $\pi_j \equiv \ket{j}\bra{j}\otimes I_d$
reflecting the symmetry of Choi-Jamiolkowski operators of
measurement $\mM$, thus, satisfying the identity
$M^T=\sum_j \pi_j M^T \pi_j$, where $T$ denotes a transposition with respect to the basis $\{\ket{j}\otimes\ket{k}\}$.
Then for any measurement
$\mM$ the conditional probabilities
satisfy the following identity
\begin{align}
p(c|\mathcal{M},\mathcal{T}) &= \tr{T_c M^T}=
\sum_j \tr{T_c\pi_j M^T \pi_j} \nonumber \\
&=\sum_j \tr{\pi_j T_c\pi_j M^T} \equiv \tr{\pi(T_c) M^T}\,,
\end{align}
implying that the test procedure formed by operators
$\{\pi(T_c)\equiv\sum_j \pi_j T_c\pi_j\}_c$
is indistinguishable from the test procedure
composed of operators $\{T_c\}_c$. In other words, without loss
of generality we may assume that the test procedure $\mathcal{T}$
is composed of positive operators of the form
\begin{align}
T_c=\sum_j \ket{j}\bra{j}\otimes H_j^{(c)}\,.
\end{align}
for which $\pi(T_c)=T_c$ and for all $j$
obeying the normalization
(see Eq.\eqref{def:norm1})
\begin{align}
\sum_c H^{(c)}_j = \rho\,.
\label{def:norm2a}
\end{align}
Consequently, the conditional probability equals
\begin{align}
 p(c|\mathcal{M},\mathcal{T}) = \sum_j \tr{H^{(c)}_j M_j}\,.
\end{align}
In the considered case of discrimination of a pair of measurements $\mM$
and $\mN$ (apriori occurring with probabilities $\eta_{\mM}$ and $\eta_{\mN}$,
respectively) we have $c\in\{\mM,\mN,f\}$, hence, the normalization
explicitly reads
\begin{align}
H^{(\mM)}_j+H^{(\mN)}_j+H^{(f)}_j = \rho\,,
\label{def:norm2}
\end{align}
for all $j=1,\dots,n$.

After deriving the above expressions we have all the mathematical instruments required to formalize and analyze any particular measurement discrimination problem. In what follows we present several cases in which the structure of the problem allows us to either partly simplify the choice of the normalization $\rho$ or to completely determine it and to reduce the optimization of the discrimination of measurements to discrimination of states.

\section{Perfect discrimination}
\label{sec:perfect}
Let us first address the case of perfect discrimination. This is an intriguing quantum information theory question, because the maximal number of simultaneously perfectly distinguishable measurements reveals the information "capacity" of measurement devices. It is known that for states this number coincides with the dimension of the Hilbert space, thus, provides its operational meaning. There are $d^2$ perfectly distinguishable unitary channels (e.g. Pauli operators in case of qubit) and this property is exploited in superdense coding \cite{dcoding} to double the information transmission rate of noiseless communication with $d$-level systems.


Surprisingly, a nontrivial insight on perfect discrimination comes
from the results of Ref.~\cite{boundariness}, where the concept of
boundariness was introduced. Based on the close relation between
boundariness and minimum-error discrimination we know that perfect
discrimination is possible only between boundary elements
(for details see section IV of \cite{boundariness}). Let us stress
that this feature holds also for states and channels.
In particular, the results of Ref.~\cite{boundariness} imply that for each
measurement $\mM$ from the boundary there exist a  measurement $\mN$
(also belonging to the boundary) such that $\mM$ and $\mN$ are perfectly
distinguishable.

The channel representation of measurements makes the problem of discrimination of observables a special case of the channel discrimination. It follows \cite{chdiscrim} that for the minimum-error discrimination of equiprobable measurements $\mM$ and $\mN$ the optimal error probability is given by the following formula
\begin{align}
\label{def:cbnorm}
p_e= \frac{1}{2}(1-\frac{1}{2}\| \mathcal{M}- \mathcal{N}\|_{\rm cb}),
\end{align}
where $\|.\|_{\rm cb}$ denotes the completely bounded (CB) norm \cite{cbnorm}.
In general, it is difficult to evaluate this norm, because
it requires inspection of the behaviour of the map when tensorized with
identity channel $\mathcal{I}_k$ on $k$ dimensional Hilbert space
$\hilb{H}_k$.
\begin{align}
\nonumber
\|\mathcal{M}-\mathcal{N}\|_{\rm cb}
=\max_{k\in \mathbb{N} ,\rho\geq 0, \tr{\rho}=1}\|[(\mathcal{M-N})\otimes \mathcal{I}_k](\rho)\|_{\rm tr},
\end{align}
where $\|X\|_{\rm tr}={\rm tr}|X|$ denotes the trace norm. Unfortunately,
the following example demonstrates that although the measurements represent
a special type of channels (with classical outputs), the perfect discrimination can not be, in general, restricted to simple (ancilla-free) schemes only, i.e.
\begin{align}
\label{eq:evalcbnorm}
\|\mathcal{M}-\mathcal{N}\|_{\rm cb} >
\max_{\rho\geq 0, \tr{\rho}=1}\|\mathcal{M}(\rho)-\mathcal{N}(\rho)\|_{\rm tr}\,.
\end{align}

\noindent{\bf Example 1} (\emph{Perfect discrimination without simple scheme}).  
\label{lem:perfect2}
Let us consider a pair of symmetric three-outcomes qubit measurements
\begin{align}
\label{eq:twotrinesog}
\mM:\ M_1&=\frac{2}{3} \ket{0}\bra{0} \,, M_2=\frac{2}{3} \ket{v_+}\bra{v_+} \,, M_3=\frac{2}{3} \ket{v_-}\bra{v_-}\,, \nonumber \\
\mN:\ N_1&=\frac{2}{3} \ket{1}\bra{1} \,, N_2=\frac{2}{3} \ket{v^\perp_+}\bra{v^\perp_+} \,, N_3=\frac{2}{3} \ket{v^\perp_-}\bra{v^\perp_-}\,,
\end{align}
where $\ket{v_\pm}=\frac{1}{2}\ket{0} \pm \frac{\sqrt{3}}{2}\ket{1}$, $\ket{v^\perp_{\,\pm}}= \frac{\sqrt{3}}{2}\ket{0}\mp \frac{1}{2}\ket{1}$.
Applying these measurements on one part of a singlet state $\ket{\psi_-}=(\ket{01}-\ket{10})/\sqrt{2}$ of two qubits the other part (ancilla) is projected into two orthogonal states for any of the (equiprobable) outcomes $j$.
For example, outcome $j=2$ heralds the ancilla state $\ket{v^\perp_+}$ in case of measurement $\mM$ and state $\ket{v_+}$ in case of $\mN$. Thus, for $j=2$ the perfect discrimination can be achieved by distinguishing orthogonal states $\ket{v_+}$, $\ket{v^\perp_+}$. Similarly, for $j=1$ ($j=3$) we would have states $\ket{0}$, $\ket{1}$ ($\ket{v_-}$, $\ket{v^\perp_-}$), respectively. We conclude that measurements $\mM$, $\mN$ can be perfectly discriminated using general (ancilla-assisted) scheme. It remains to show there is no ancilla-free scheme for perfect discrimination. Let us denote by $\mu_j=\tr{M_j\varrho}$ and
$\nu_j=\tr{N_j\varrho}$ the probabilities of outcomes $j$ given the probe
state is $\varrho$.
For ancilla-free scheme the perfect discrimination happens if and only if each outcome $j$ is associated either with conclusion $\mathcal{M}$
or $\mathcal{N}$ and on top of that the probability of each outcome is non-vanishing for at most one of the measurements, i.e. $\sum_j \mu_j\nu_j=0$ (being equivalent to the conditions $\mu_j\nu_j=0$ for
each $j$).
For the considered pair of measurements it follows that
always at least two of the outcomes have nonvanishing probabilities, thus,
the necessary condition 
for perfect discrimination ($\sum_j \mu_j \nu_j=0$) can not be satisfied.
In conclusion, for the discrimination of measurements the use of ancilla
provides a nontrivial advantage in comparison with simple schemes.

\subsection{Binary measurements}
In this section we will focus on perfect discrimination of
two outcomes (binary) measurements. Suppose
measurements $\mathcal{M}$, $\mathcal{N}$
are described by effects $M_1,M_2$ ($M_1+M_2=I$)
and $N_1,N_2$ ($N_1+N_2=I$), respectively. The following
theorem provides a simple criterion for binary measurements
being perfectly distinguishable. Moreover, it
justifies that their perfect discrimination
is achievable by simple schemes. 

\begin{theorem}
\label{lem:perfect4}
A pair of two outcome measurements $\mathcal{M}$ and $\mathcal{N}$
can be perfectly discriminated if and only if
there exist a state $\ket{\psi}$
such that
\begin{align}
\bra{\psi}M_j\ket{\psi}&=1 \quad {\rm and}
\quad \bra{\psi}N_j\ket{\psi}=0\,,
\label{eq:binperfcond}
\end{align}
for either $j=1$, or $j=2$.
\end{theorem}

\Proof
We start by proving sufficiency of the condition.
Suppose $j=1$, i.e. the identities
$\mu_1=\bra{\psi}M_1\ket{\psi}=1$ and $\nu_1=\bra{\psi}N_1\ket{\psi}=0$ hold.
The normalization implies $\mu_2=0$ and $\nu_2=1$, hence,
applying the unknown measurement on the probe state $\ket{\psi}$
and recording the outcome $1$ we may conclude with certainty
that the measurement is $\mathcal{M}$. Similarly, the observation
of the outcome $2$ implies the unknown measurement is $\mathcal{N}$,
thus, the perfect discrimination is achieved. The argumentation for
the case $j=2$ is analogous, 
only interpretation of the observed
outcomes is switched. This proves the sufficiency of the 
identities (\ref{eq:binperfcond}).

Let us proceed and prove their necessity. First we will show that
perfect discrimination conditions for binary measurements $\mathcal{M}$
and $\mathcal{N}$
\begin{align}
0=p(\mN|\mM,\mathcal{T})=\tr{H^{(\mN)}_1 M_1}+\tr{H^{(\mN)}_2 M_2}\,, \nonumber \\
0=p(\mM|\mN,\mathcal{T})=\tr{H^{(\mM)}_1 N_1}+ \tr{H^{(\mM)}_2 N_2}\,,
\label{eq:zero}
\end{align}
implies $\tr{\varrho(M_1+N_1)}=1$. Since the trace of a product of two
positive operators is nonnegative all four traces in the above equation
vanish. In particular, the condition $\tr{H_2^{(\mN)}M_2}=0$ implies
$\tr{H_2^{(\mN)}M_1}=\tr{H_2^{(\mN)}(I-M_2)}=\tr{H_2^{(\mN)}}$. Similarly,
$\tr{H_2^{(\mM)}N_2}=0$ implies $\tr{H_2^{(\mM)}N_1}=\tr{H_2^{(\mM)}}$.
Using the identity 
\begin{align}
\label{eq:normbinary}
H^{(\mM)}_2+H^{(\mN)}_2=\rho
\end{align}
implied by normalization (\ref{def:norm2a}) with $\tr{\rho}=1$
we obtain the condition
\begin{align}
\tr{H^{(\mN)}_2 M_1}+\tr{H^{(\mM)}_2 N_1}=1\, .   
\label{eq:sumone}
\end{align}
Due to Eq. (\ref{eq:normbinary}) we have $H^{(\mN)}_2\leq \rho$ and $H^{(\mM)}_2\leq\rho$, hence
\begin{align}
\label{eq:lbound}
1=\tr{H^{(\mN)}_2 M_1}+\tr{H^{(\mM)}_2 N_1}\leq\tr{\varrho(M_1+N_1)}\,.
\end{align}

Trace of a product of two positive operators vanishes if and
only if the supports of the two operators are orthogonal.
This implies (Eqs.\eqref{eq:zero}, (\ref{def:norm2a}))
\begin{align}
\label{def:mtilda}
H^{(\mN)}_1&=\lambda \, \widetilde{M}_1^\perp\,,  \quad\quad
H^{(\mM)}_1=(1-\lambda) \widetilde{N}_1^\perp \, ;
\end{align}
where $\widetilde{M}_1^\perp$, $\widetilde{N}_1^\perp$ are density operators with
supports orthogonal to $M_1$, $N_1$, respectively, and
$\lambda=\tr{H^{(\mN)}_1}\in[0,1]$. Clearly due to normalization
$$
\rho=H_1^{(\mM)}+H_1^{(\mN)}=\lambda \widetilde{M}_1^\perp
+(1-\lambda)\widetilde{N}_1^\perp\,.
$$
Using the identities $\widetilde{M}_1^\perp M_1=O$ and
$\widetilde{N}_1^\perp N_1=O$ we obtain
\begin{align}\nonumber
\tr{\rho(M_1+N_1)}&=
\lambda\tr{\widetilde{N}_1^\perp M_1}+(1-\lambda)\tr{\widetilde{M}_1^\perp N_1}\\
&\leq  \max\{\tr{\widetilde{N}_1^\perp M_1},\tr{\widetilde{M}_1^\perp N_1}\}\,.
\label{eq:upboundgamma}
\end{align}
Since $\widetilde{M}_1^\perp$, $\widetilde{N}_1^\perp$ are states and $M_1$, $N_1$
are effects (i.e. $M_1,N_1 \leq I$) it follows that
$\tr{\rho(M_1+N_1)}\leq 1$. Combining this inequality with
Eq.~\eqref{eq:lbound} we may conclude that perfect discrimination
implies $\tr{\varrho(M_1+N_1)}=1$,
thus, the upper and lower bounds are both saturated. For upper bound
this requires an existence either of a state $\rho=\widetilde{M}_1^\perp$
such that $\tr{\widetilde{M}_1^\perp N_1}=1$, $\tr{\widetilde{M}_1^\perp M_1}=0$,
or of a state $\rho=\widetilde{N}_1^\perp$ such that
$\tr{\widetilde{N}_1^\perp M_1}=1$, $\tr{\widetilde{N}_1^\perp N_1}=0$.
Finally, let us stress that the state $\rho$ can be always
chosen to be a pure state $\ket{\psi}$ being an eigenvector of
$M_1$ (case $j=1$), or $N_1$ (case $j=2$) associated with eigenvalue $1$
and simultaneously belonging to the kernel of operators $N_1$, $M_1$,
respectively.
\qed

As a consequence of this theorem a pair of $2$-outcome (binary) measurements can be perfectly discriminated only if one of the POVM elements of $\mM$ (say $M_1$) has eigenvalue one in a subspace in which $N_1$ has eigenvalue zero, i.e. $M_1\ket{\psi}=\ket{\psi}$ and $N_1\ket{\psi}=0$. In particular, for binary qubit measurements the perfect distinguishability implies the following form of observables
\begin{align}
M_1=\ket{\varphi}\bra{\varphi}+ q \ket{\varphi^\perp}\bra{\varphi^\perp} \quad M_2=(1-q)\ket{\varphi^\perp}\bra{\varphi^\perp} \nonumber \\
N_1= r \ket{\varphi^\perp}\bra{\varphi^\perp} \quad N_2=\ket{\varphi}\bra{\varphi}+(1-r)\ket{\varphi^\perp}\bra{\varphi^\perp}, \nonumber
\end{align}
where $0\leq r,q\leq 1$ and $\ket{\varphi}$, $\ket{\varphi^\perp}$ form an orthonormal basis of $\hilb{H}_2$. The optimal probe state reads
$\ket{\psi}=\ket{\varphi}$.

\subsection{More than two measurements}
Let us get back to the questions raised at the beginning of this section.
What is the maximum number $m$ of perfectly distinguishable measurements
of $d-$dimensional quantum system? How is this number related to the
dimension? In what follows we
give an example exhibiting a rather
surprising fact that $m$ can be arbitrary, irrelevant of the system's
dimension.

Consider 
$m$ measurements $\mM_l$ ($l=1,\dots,m$),
each of them with $n\geq m$ outcomes (labeled as before by $j=1,\dots,n$)
associated with effects
\begin{align}
M_{lj} =
\left\{\begin{array}{lcl}
\ket{\varphi}\bra{\varphi} & {\rm if}& j=l \\
x_{lj}(I-\ket{\varphi}\bra{\varphi}) & {\rm if}& j\neq l
\end{array}
\right. , \nonumber
\end{align}
where $0<x_{lj}<1$ and $\sum_j x_{lj}=1$. Using a test state
$\ket{\psi}=\ket{\varphi}$ the outcome $j=l$ of measurement
$\mM_l$ is observed with certainty, hence, observation
of the outcome $j$ perfectly identifies the measurement
$\mM_{l=j}$. This is an example of $m$ perfectly distinguishable
measurements. Let us stress that the dimension of the system
is not specified and also that no ancilla is needed. Let us also
note that the choice of $x_{lj}$ (for $j\neq l$) is arbitrary, thus,
the measurements $\mM_l$ are not just mutually relabeled
measurements.

The following proposition relates the maximal
number of perfectly distinguishable measurements $m$
with the number of outcomes $n$.

\begin{proposition}
\label{lem:perfect3}
If $n-$outcome
quantum measurements $\mM_1,\dots,\mM_m$ can be perfectly
discriminated then $m \leq n$.
\end{proposition}

\Proof
Similarly, as for the discrimination of two measurements we can write the conditional probability as  $p(c|\mathcal{M}_l,\mathcal{T})=\sum_j \tr{H^{(c)}_j M_{lj}}$. Since for perfect discrimination the inconclusive outcome cannot occur, thus, $c\neq f$, we
use $c\in\{1,\dots,m\}$ indicating the measurement $\mM_c$.
The operators $H^{(c)}_j$ must fulfill the normalization identity
$\sum_{k=1}^m \; H^{(k)}_j = \rho$
for $\forall j\in \Omega$.
A test $\mathcal{T}$ perfectly distinguishes measurements $\{\mathcal{M}_l\}$
if and only if for all $l$ the following identity holds
$p(c|\mathcal{M}_l,\mathcal{T})=\sum_j \tr{H^{(c)}_j M_{lj}}=\delta_{cl}$.

Let us introduce positive operators
$E_{cj} \equiv \rho^{-1/2} H^{(c)}_j \rho^{-1/2}$ and $Q_{lj} \equiv \rho^{1/2} M_{lj} \,\rho^{1/2}$
satisfying the identities
\begin{align}
\sum_{c} E_{cj} = \Pi_{\rho} \quad{\rm and}\quad \sum_{j=1}^n Q_{lj} = \rho
\label{def:sumeq}
\end{align}
for all $j$ and $l$, respectively. We denoted by $\Pi_{\varrho}$
the projector onto the support of $\varrho$.
Then
\begin{align}
p(l|\mM_l,\mathcal{T})&=\sum_j \tr{E_{lj} Q_{lj}}\\
\nonumber
&\leq\sum_j
\tr{\Pi_{\rho} Q_{lj}}=\sum_j \tr{Q_{lj}}=1\,,
\end{align}
where we used that $E_{lj}\leq \Pi_{\rho}$,
 $Q_{lj}\leq \rho\leq \Pi_{\rho}$,
Eq. \eqref{def:sumeq} and $\tr{\rho}=1$.
It follows that the condition
$p(l|\mM_l,\mathcal{T})=1$
holds only if $\tr{E_{lj} Q_{lj}}=\tr{Q_{lj}}$
for all $l,j$. Since $0\leq E_{lj}\leq \Pi_{\rho}$ and $0 \leq Q_{lj}\leq \Pi_{\rho}$ this is equivalent to
the requirement $E_{lj} \geq \Pi_{lj}$, where $\Pi_{lj}$ denotes a projector
onto a support of the operator $Q_{lj}$. Consequently, the multiplicity
$\kappa_{lj}$ of eigenvalue $1$ in the spectral decomposition of $E_{lj}$
has to be at least the rank of $\Pi_{lj}$, i.e.
\begin{align}
\label{eq:ineqkappalj}
\forall l,j \quad \kappa_{lj}\geq \tr{\Pi_{lj}}.
\end{align}
Denote by $D=\tr{\Pi_{\varrho}}$
the dimension of the support of $\rho$.
Due to Eq. (\ref{def:sumeq}) we have
$\sum_j \tr{\Pi_{lj}}\geq D$,
because the rank of the sum of positive operators $Q_{lj}$ is at most the sum
of the ranks of its parts.
Combining this with Eq. (\ref{eq:ineqkappalj}) and summing over $l$ we obtain
\begin{align}
\sum_l\sum_j \kappa_{lj} \geq m D.
\label{eq:sumrulemu1}
\end{align}
On the other hand, taking into the account the normalization
from Eq.~\eqref{def:sumeq} and inequality $\kappa_{lj}\leq \tr{E_{lj}}$
it follows that  $\sum_l \kappa_{lj} \leq D$ and consequently
\begin{align}
\sum_j\sum_l\kappa_{lj} \leq n D\,.
\label{eq:sumrulemu2}
\end{align}
Combining inequalities (\ref{eq:sumrulemu1}), and (\ref{eq:sumrulemu2}) we get $m \leq n$.
\qed

For non-degenerate projective measurements a rank one projector
corresponds to each outcome.
By definition such measurement has $n=d$ outcomes, where $d$ is the dimension
of $\hilb{H}_d$. The above Proposition \ref{lem:perfect3} implies there
are at most $d$ perfectly distinguishable non-degenerate
projective measurements.

\section{Quantum filters}
\label{sec:qfilters}
A projective two outcome measurement $\mM$ is called a \emph{quantum filter}
if one of its outcomes is described by rank-one projection. Discrimination
of a pair of quantum filters $\mM,\mN$ can appear in two different variations
depending on assignment of rank-one operators to particular labels: either
the same outcome is described by rank-one operators, or exclusive outcomes
are associated with rank-one operators for $\mM$ and $\mN$.

Let us start with the first case and set the outcome labeled as "1"
to be the one described by rank-one projector, i.e.
\begin{align}
\label{eq:qfilters}
 \mM &: M_1 = \ket{\varphi} \bra{\varphi}\,,  \quad  M_2=I-M_1 \,;\nonumber \\
 \mN &: N_1 = \ket{\psi} \bra{\psi}\, , \quad\  N_2=I-N_1\,.
\end{align}
The reduction theorem formulated in the following section reduces this problem
to discrimination of qubit projective measurements by identifying
a relevant two-dimensional subspace of $\hilb{H}_d$. In particular,
the statement of the theorem is more general and allows us to identify
irrelevant subspace for discrimination of arbitrary measurements.

\subsection{Reduction theorem}
\label{sec:redtheorem}
Consider a pair of $n-$outcome
measurements $\mM$ and
$\mN$ on $d-$dimensional Hilbert space represented by POVMs
$\{M_j\}$ and $\{N_j\}$, respectively.
Suppose that $\forall j$ $Q_j$ is a (largest) projector such that
$Q_j\leq M_j$ and $Q_j\leq N_j$. Due to POVM normalization
the projectors $Q_j$ are mutually orthogonal, i.e.
$Q_jQ_k=\delta_{jk}Q_j$. We may define a projector $P=\sum_j Q_j$
and "measurements" $\widetilde{\mM}$ and $\widetilde{\mN}$ with POVM elements
$\{\widetilde{M}_j\equiv M_j-Q_j\}$ and $\{\widetilde{N}_j\equiv N_j-Q_j\}$,
respectively, and normalized to $I-P$. Let us stress that
$\widetilde{M}_j=(I-P)M_j(I-P)$. The following theorem shows that the subspace
determined by the support of $P$ plays no role
and the original discrimination problem is equivalent to discrimination
of measurements $\widetilde{\mM},\widetilde{\mN}$ defined on
the subspace $\widetilde{\hilb{H}}\equiv(I-P)\hilb{H}_d$ of $\hilb{H}_d$
relevant for the discrimination.

\begin{theorem}
\label{lem:reduction1}
Suppose that $\mathcal{T}$ and $\widetilde{\mathcal{T}}$ are optimal solutions
to discrimination
with fixed failure rate $p_f=\widetilde{p}_f$
between pairs of measurements $\mM,\mN$ and
$\widetilde{\mM},\widetilde{\mN}$, respectively. Then
$p_s=\widetilde{p}_s$ or equivalently $p_e=\widetilde{p}_e$.
Moreover, optimal $\mathcal{T}$ can be chosen such that
$\mathcal{T}|_{\widetilde{\hilb{H}}}=\widetilde{\mathcal{T}}$
and vice versa (i.e. given optimal $\mathcal{T}$ optimal test $\widetilde{\mathcal{T}}$ can be chosen as
$\mathcal{T}|_{\widetilde{\hilb{H}}}=\widetilde{\mathcal{T}}$).

\end{theorem}
\Proof
See appendix \ref{sec:appreduction}.
\qed

Let us formulate consequences of the above theorem for quantum filters.
We define a two dimensional Hilbert space
$\widetilde{\hilb{H}}$ as a linear span of vectors
$\ket{\varphi}$, $\ket{\psi}$. Clearly, $M_2\geq P$ and $N_2\geq P$,
where $P$ is a projector onto the subspace $\widetilde{\hilb{H}}^\perp$.
Using the reduction theorem \ref{lem:reduction1} the discrimination
of filters can be solved by finding the solution to discrimination
of projective qubit measurements
\begin{align}
\label{eq:qfilters-reduction}
 \widetilde{\mM}:\quad \widetilde{M_1} = \ket{\varphi} \bra{\varphi}\,,  \quad  \widetilde{M_2}=\ket{\varphi^\perp}\bra{\varphi^\perp} \,;\nonumber \\
 \widetilde{\mN}:\quad \widetilde{N_1} = \ket{\psi} \bra{\psi} \,,\quad  \widetilde{N_2}=\ket{\psi^\perp}\bra{\psi^\perp}\,,
\end{align}
where $\ket{\psi^\perp},\ket{\varphi^\perp}$ are vectors from $\widetilde{\hilb{H}}$ orthogonal to $\ket{\psi},\ket{\varphi}$, respectively. The solution to this problem is given in Section \ref{sec:projqm}. Let us stress that similar reasoning applies also to the case of discrimination among $m$ quantum filters (with $M_{l1}$ being rank-one projectors). In such case, the problem is equivalent to discrimination of $m$ quantum filters on $m$-dimensional subspace of $\hilb{H}_d$.

Finally, we discuss the other possible assignment of outcomes for two quantum filters, i.e. the case
\begin{align}
\label{eq:qfilters1}
 \mM &: M_1 = \ket{\varphi} \bra{\varphi}\,,  \quad  M_2=I-M_1 \,;\nonumber \\
 \mN &: N_1 = I-N_2\, , \quad\  N_2=\ket{\psi} \bra{\psi}\, ,
\end{align}
when rank one projections correspond to different outcomes.
If the dimension of $\hilb{H}_d$ is two the problem coincides with discrimination of two projective qubit measurements, which we solve in the next section.
Otherwise, there exists a state $\ket{\phi}$ orthogonal to $\ket{\psi}$ and $\ket{\varphi}$.
Measuring $\ket{\phi}$ with $\mM$ we always get outcome $2$, while $\mN$ will always produce outcome $1$. Thus, for $\dim\hilb{H}_d \geq 3$ any pair of (different) quantum filters (\ref{eq:qfilters1}) is always perfectly distinguishable.

\section{Projective qubit measurements}
\label{sec:projqm}
In this section we shall analyze discrimination of projective qubit measurements, i.e. measurements such as
$\mM$ described by effects $M_1 = \ket{\varphi} \bra{\varphi}$ and $M_2=I-M_1= \ket{\varphi^\perp} \bra{\varphi^\perp}$ for some orthonormal basis $\{\ket{\varphi},\ket{\varphi^\perp}\}$ of $\hilb{H}_2$. As we declared in section \ref{sec:mft} our goal is to maximize probability of success $p_s$ for a fixed failure probability $p_f$.

\subsection{Binary discrimination problem}
\label{sec:binary_disc_problem}
Let us start with the simplest case, when our goal is to discriminate among pair of projective measurements
\begin{align}
& \mM:\quad M_1 = \ket{\varphi} \bra{\varphi}\,, \quad & M_2= \ket{\varphi^\perp} \bra{\varphi^\perp}\,; \nonumber \\
& \mN:\quad N_1 = \ket{\psi} \bra{\psi}\,, \quad & N_2=\ket{\psi^\perp} \bra{\psi^\perp}\,.
\label{def:perp}
\end{align}
Suppose $\mathcal{T}$ is a test procedure
specified by operators $H^{(c)}_1, H^{(c)}_2$ with $c \in \{\mM,\mN,f\}$ such that for all $j$ $\sum_c H_j^{(c)}=\varrho$ and
$\mathcal{T}$ leads to certain values of $p_s,p_e$ and $p_f$.
Further, we will exploit the reflection symmetry of the problem. In particular, let us denote by $\Gamma$ the \emph{universal NOT} transformation $X\mapsto X^\perp=\tr{X}I-X$ for any operator $X$.
In $\hilb{H}_2$ this map is positive (not completely positive) and trace-preserving. Moreover, $\tr{\Gamma(X)Y}=\tr{X\Gamma(Y)}$ for all operators $X,Y$ and $\Gamma^2=\mathcal{I}$.

By properties (positivity) of $\Gamma$ it follows that operators
\begin{align}
& H'^{(c)}_1 = \Gamma(H^{(c)}_2)\;, \quad {\rm and}\quad
 H'^{(c)}_2 = \Gamma(H^{(c)}_1)\;,
\label{def:tprime}
\end{align}
form a valid test procedure $\mathcal{T}^\prime$ with normalization
\begin{align}
\sum_c H_1^{\prime (c)} =\sum_c H_2^{\prime (c)} =\Gamma(\rho)\;. \nonumber
\end{align}
For conditional probabilities we find
\begin{align}
\label{eq:condpr3}
p(c|\mathcal{M},\mathcal{T'}) & = \tr{\Gamma(H^{(c)}_2)\; \ket{\varphi}\bra{\varphi} + \Gamma(H^{(c)}_1)\; \ket{\varphi^\perp}\bra{\varphi^\perp}}\nonumber \\
& = \tr{H^{(c)}_2 \; \Gamma(\ket{\varphi}\bra{\varphi}) + H^{(c)}_1 \; \Gamma(\ket{\varphi^\perp}\bra{\varphi^\perp})} \nonumber \\
& = \tr{H^{(c)}_2 \ket{\varphi^\perp}\bra{\varphi^\perp} + H^{(c)}_1 \ket{\varphi}\bra{\varphi}} \nonumber \\
& = p(c|\mathcal{M},\mathcal{T})
\end{align}
and analogously $p(c|\mathcal{N},\mathcal{T'})=p(c|\mathcal{N},\mathcal{T})$.
In other words, both test procedure $\mathcal{T}$ and $\mathcal{T}^\prime$
determine the same probabilities $p_s$, $p_e$ and $p_f$, thus they both perform
equally well in the considered
discrimination problem. Moreover, any convex combination, in particular
$\widetilde{\mathcal{T}}=\frac{1}{2}\mathcal{T}+\frac{1}{2}\mathcal{T}^\prime$ of
these tests, results in the same
probabilities $p_s$, $p_e$ and $p_f$.
This allows us to reduce the set of considered test procedures and to fix their normalization $\rho$ without loss of generality,
because the normalization of $\widetilde{\mathcal{T}}$ is independent of the test $\mathcal{T}$
and reads
\begin{align}
\label{eq:normondcov1}
\forall j \quad\sum_c \widetilde{H}_j^{(c)}&= \sum_c \frac{1}{2}(H_j^{(c)}+H_j^{\prime (c)}) \nonumber \\
&=\frac{1}{2}\Big(\varrho+\Gamma(\varrho)\Big)=\frac{1}{2}I\,.
\end{align}
Moreover,
\begin{align}
\Gamma(\widetilde{H}_1^{(c)})&=\Gamma\Big(\frac{1}{2}(H_1^{(c)}+H_1^{\prime (c)})\Big) \nonumber \\
&=\frac{1}{2}(H_2^{\prime (c)}+H_2^{(c)})=\widetilde{H}_2^{(c)}\, ,
\end{align}
so the considered test procedures $\widetilde{\mathcal{T}}$ are completely specified by operators
for a single outcome, i.e. by positive operators $\widetilde{H}_1^{(c)}$ and by their normalization condition (\ref{eq:normondcov1}).
Using this fact
we obtain formulas
\begin{align}
\label{eq:condprob3}
 p(c|\mathcal{M},\widetilde{\mathcal{T}}) &=  \tr{\widetilde{H}^{(c)}_1 \ket{\varphi}\bra{\varphi} + \Gamma(\widetilde{H}^{(c)}_1) \ket{\varphi^\perp}\bra{\varphi^\perp}} \nonumber \\
 & = \tr{2\widetilde{H}^{(c)}_1 \ket{\varphi}\bra{\varphi}} \equiv \tr{E_c \ket{\varphi}\bra{\varphi}} \nonumber \\
 p(c|\mathcal{N},\widetilde{\mathcal{T}}) & =\tr{2\widetilde{H}^{(c)}_1 \ket{\psi}\bra{\psi}} \equiv \tr{E_c \ket{\psi}\bra{\psi}}\,,
\end{align}
where we defined positive operators $E_c \equiv 2\widetilde{H}^{(c)}_1$ for each
$c\in\{\mM,\mN,f\}$. Let us stress that Eq.~\eqref{eq:normondcov1} implies
$$
E_{\mM}+E_{\mN}+E_{f}=I\, .
$$
In other words, the positive operators $E_{\mM},E_{\mN},E_{f}$ form a POVM coinciding with a measurement discriminating pure states $\ket{\psi},\ket{\varphi}$.
Indeed, using Eqs. (\ref{def:q}), (\ref{def:pepc}), (\ref{eq:condprob3}) we can express $p_s,p_e$ and $p_f$ as
\begin{align}
p_s&= \eta_\mathcal{M}\; \bra{\varphi}E_{\mM} \ket{\varphi} +\eta_\mathcal{N}\; \bra{\psi} E_{\mN} \ket{\psi} \nonumber\\
p_e&= \eta_\mathcal{M}\; \bra{\varphi}E_{\mN} \ket{\varphi} +\eta_\mathcal{N}\; \bra{\psi} E_{\mM} \ket{\psi} \nonumber\\
p_f&= \eta_\mathcal{M}\; \bra{\varphi}E_f \ket{\varphi} +\eta_\mathcal{N}\; \bra{\psi} E_f \ket{\psi} \, .
\end{align}
Thus, we managed to reduce the discrimination of projective qubit measurements (in any version) to discrimination of pure states. In particular, we may formulate the following theorem.

\begin{theorem}
The problem of optimal discrimination with fixed failure rate $p_f$ of projective qubit measurements $\mM$ and $\mN$ (determined by vector states $\ket{\varphi},\ket{\psi}$, respectively)  is mathematically equivalent to an optimal discrimination with fixed failure rate $p_f$ of pure states $\ket{\varphi}$, $\ket{\psi}$.
\end{theorem}

Suppose a POVM associated with effects $E_c$ with $c\in\{\varphi,\psi,f\}$ is the optimal solution (for details see \cite{intmdt0,intmdteq1,intmdt2})
for the discrimination with a fixed failure rate $p_f$ of pure states $\ket{\varphi},\ket{\psi}$. Then the optimal discrimination (see Eq.~\eqref{def:perp}) of projective qubit measurements $\mM$ and $\mN$ (determined by vector states $\ket{\varphi},\ket{\psi}$, respectively) can be implemented as follows. We prepare a maximally entangled state $\ket{\phi}=(\ket{00}+\ket{11})/\sqrt{2}$ of two qubits. We measure one of the qubits by the unknown measurement that we want to identify. If the outcome $1$ is observed, then we perform the measurement of $\{E_{c}^T\}$ on the second qubit. If we observe outcome $2$, then the second qubit is measured by POVM $\{\Gamma(E_{c})^T=\Gamma(E^T_{c})\}$. It is straightforward to verify that this procedure results in conditional probabilities given in Eq.~\eqref{eq:condprob3} if we associate conclusions $c$ as $\varphi \leftrightarrow \mM$, $\psi \leftrightarrow \mN$.
Let us remind that the reduction theorem \ref{lem:reduction1} described in the previous section implies that the same procedure can be used to discriminate (optimally) quantum filters.

In the following we illustrate what the results on optimal discrimination of two states imply for the discrimination of two qubit measurements.

\noindent{\bf Example 2} (\emph{Minimum error discrimination}).
By definition we set $p_f=0$. The formula for discrimination of two pure states
is well-known due to seminal works of Helstrom and Holevo \cite{holevo,helstrom}. Let us denote by
$\eta\equiv \eta_\mM$ the apriori probability for $\ket{\varphi}$ ($\mM$) and $1-\eta\equiv \eta_\mN$
being the apriori probability for $\ket{\psi}$ ($\mN$). The optimal POVM
consists of elements $E_\varphi=\ket{\alpha}\bra{\alpha}$,
$E_\psi=\ket{\beta}\bra{\beta}$ being projectors onto positive and negative
eigensubspaces of operator $\Delta=(1-\eta)\ket{\psi}\bra{\psi} - \eta\ket{\varphi}\bra{\varphi}$, respectively. The optimal (minimal) probability of error is given by the famous Helstrom's formula
\begin{align}
p_e=\frac{1}{2}(1-\sqrt{1-4\eta(1-\eta) |\bra{\psi} \varphi \> |^2})\,.
\end{align}
The test procedure described above helps us to design the optimal discrimination of pair of associated measurements $\mM$ and $\mN$. However, let us note that the ancilla is not really necessary to achieve the optimality (which is in accordance with the discussion at the beginning of Section \ref{sec:perfect}). Indeed, it is sufficient to prepare a test state $\ket{\alpha}$. Observing outcome $1$ we conclude that the tested measurement was $\mN$ and otherwise we conclude it was $\mM$. Alternatively, one can also exploit the test state $\ket{\beta}$ and inverting the interpretation of the outcomes, we achieve again the optimal value of the error probability $p_e$.

\noindent{\bf Example 3} (\emph{Unambiguous discrimination}).
By definition we call the discrimination unambiguous if $p_e=0$. The solution to unambiguous discrimination of two pure states
was found first for equal prior probabilities by Ivanovic \cite{idp1}, Dieks \cite{idp2} and Peres \cite{idp3} and later by Jaeger and Shimony \cite{idp4} for the general situation. The solution has three regimes depending on the relation between the prior probability $\eta$ and overlap  $F=|\<\psi|\varphi\>|$
\begin{equation}
\label{eq:pfunamb}
p_f=\left\{\begin{array}{ll}
\eta+(1-\eta)F^2 & \quad (1+F^2)\eta \leq F^2\,; \\
2\sqrt{\eta(1-\eta)} F & \quad F^2\leq (1+F^2)\eta\leq 1\,; \\
1-\eta+\eta F^2 & \quad (1+F^2)\eta \geq 1 \,.
\end{array}\right.
\end{equation}
If the priors are very unbalanced (first and last intervals) then one of the states is never detected, so the optimal measurement has two outcomes and is projective. In the intermediate regime when the priors are "comparable" all three outcomes have non zero probability of appearance.

This means that also for discrimination of projective qubit measurements we will have three regimes defined by the same conditions. For the regime of "comparable" prior probabilities it is clear that we need to use the ancillary measurement, since we need three outcomes. An intuitive scheme for achieving the optimal performance is based on preparing a singlet state $(\ket{01}-\ket{10})/\sqrt{2}$ of two qubits. Application of the unknown measurement on one part of the state projects (depending on the identity of the measurement) the other part into a state $\ket{\varphi^\perp}$ or $\ket{\psi^\perp}$ in case of outcome $1$ and into state $\ket{\varphi}$ or $\ket{\psi}$ in case of outcome $2$. These two pairs of states have the same overlap ($|\bra{\varphi} \psi \>|=|\bra{\varphi^\perp} \psi^\perp \>|=F$) and we can discriminate within the pairs using the optimal unambiguous pure state discrimination by Jaeger and Shimony. Thanks to equal overlap in case of outcome $1$, outcome $2$ and also on average we fail with the probability $p_f$ given in Eq. (\ref{eq:pfunamb}).

For the remaining (unbalanced) regimes
the optimal performance can be achieved also by directly measuring the single partite state with the unknown measurement. If $(1+F^2)\eta_\mathcal{M}\geq 1$ then we prepare $\ket{\psi^\perp}$ and the outcome $1$ unambiguously indicates that the unknown measurement is $\mM$, whereas the outcome $2$ is inconclusive and means that the test failed. Similarly, if $(1+F^2)\eta_\mathcal{M}\leq F^2$, then we use $\ket{\varphi^\perp}$ as the test state and outcome $1$ unambiguously identifies the measurement $\mN$.

\noindent{\bf Example 4} (\emph{Noisy qubit measurements})
\label{sec:noisyqubitm}
Suppose $\mM, \mN$ are defined as convex combinations of a projective measurement and a trivial observable generating the uniform distribution of outcomes independently of the measured state, i.e.
\begin{align}
& M_1 = \mu \ket{\varphi} \bra{\varphi} +\frac{1-\mu}{2}I  \quad & M_2=\mu \ket{\varphi^\perp} \bra{\varphi^\perp} +\frac{1-\mu}{2}I \nonumber \\
& N_1 = \nu \ket{\psi} \bra{\psi} +\frac{1-\nu}{2}I \quad & N_2=\nu \ket{\psi^\perp} \bra{\psi^\perp} +\frac{1-\nu}{2}I\,.
\label{def:noisym}
\end{align}
As the key symmetry $\Gamma(M_1)=M_2$, $\Gamma(N_1)=N_2$ holds, we can directly generalize the arguments used before and conclude that the optimal test procedure is characterized by POVM elements $E_{\mM}, E_{\mN}, E_f$, which thanks to this symmetry define
$\widetilde{H}^{(c)}_1=\frac{1}{2}E_c$, $\widetilde{H}^{(c)}_2=\frac{1}{2}\Gamma(E_c)$. We find
\begin{align}
p_s= \eta_\mathcal{M}\; \tr{E_{\mM} M_1} +\eta_\mathcal{N}\; \tr{E_{\mN} N_1}\,, \nonumber\\
p_e= \eta_\mathcal{M}\; \tr{E_{\mN} M_1} +\eta_\mathcal{N}\; \tr{E_{\mM} N_1}\,, \nonumber\\
p_f= \eta_\mathcal{M}\; \tr{E_f M_1} +\eta_\mathcal{N}\; \tr{E_f N_1}\,. \nonumber
\end{align}
Let us stress that operators $M_1, N_1$ are positive and have trace one, so they correspond to mixed quantum states. Thus, for measurements $\mM$, $\mN$ defined by POVM elements from Eq. (\ref{def:noisym}) we re-expressed the problem as discrimination with fixed failure rate $p_f$ among two mixed states $M_1$, $N_1$. Such problems were studied in \cite{intmdt1}
and an upper bound on the success probability was derived. Notice that for $\mu\nu\neq 0$ the unambiguous discrimination of measurements is not possible, because the states $M_1,N_1$ have completely overlapping supports \cite{overlapsupp}. For the minimum error discrimination the optimal error rate and optimal POVM $\{E_c\}$ can be again acquired easily from the work of Helstrom \cite{helstrom}.

\subsection{General case}
\label{sec:dmm}
\label{sec:multipleffr}

In general, the discrimination of more than two objects is more complicated than the discrimination among two of them. However, in our particular case, it turns out that the derivations in the section \ref{sec:projqm} can be trivially generalized to the discrimination of $m$ projective qubit measurements. In particular, the optimal discrimination of measurements $\mM_1,\dots,\mM_m$, each of them associated with effects $M_{l1}=\ket{\varphi_l}\bra{\varphi_l}$ and $M_{l2}=\ket{\varphi^\perp_l}\bra{\varphi^\perp_l}$, can be designed by using the optimal measurement discriminating the pure states $\ket{\varphi_1},\dots,\ket{\varphi_m}$.
The optimal performance can be achieved by preparing a maximally entangled state, measuring one part of it by the unknown measurement and optimally discriminating the states of the remaining system.
Thus, the optimal relation between success $p_s$ and failure probability $p_f$ for the discrimination of projective qubit measurement apparatuses is given by the solution of (pure) state discrimination problem.
Sugimoto et. al. \cite{sugimoto} solved the discrimination with fixed failure rate of three symmetric states of a qubit, but in general the solution is not known.
The special case of minimum error discrimination of $m$ pure qubit states can be solved completely using the result of Ref.~\cite{genqubitme}.

As we discussed already in section \ref{sec:perfect} the optimal discrimination of two quantum measurements with minimum possible error can always be realized by a simple discrimination scheme. A natural question arises whether a simple schemes can be utilized to perform the optimal minimum error discrimination of $m$ quantum measurements. The following example demonstrates that this is not the case, so also for projective qubit measurements there are situations when ancilla-assisted scheme is necessary for optimization of minimum-error discrimination.

\noindent
{\bf Example 5} (\emph{Minimum-error discrimination of $3$ projective qubit measurements}.)
Consider projective qubit measurements $\mM_1,\mM_2,\mM_3$
determined by states
$\ket{0}$, $\ket{v_\pm}=\frac{1}{2}(\ket{0} \pm \sqrt{3}\ket{1})$,
respectively, appearing with equal prior probabilities. Due to result of
Clarke et. al. \cite{clarke}
we have the minimal error probability
for discrimination of these states having pairwise the same fidelity. In this particular case
$p_e^{\rm opt}=1/3$, hence $p_s^{\rm opt}=2/3$ and the same holds for optimal minimum error discrimination of measurements $\mM_1,\mM_2,\mM_3$.
Let us denote by $\varrho$ the (ancilla-free) test state and
we define $x_l\equiv \tr{M_{l1} \rho}$. We further denote by $q(l|j)$, the conditional probability of conclusion $l$
given the outcome $j$ was recorded on the unknown measurement we would
like to identify. By definition $\sum_l q(l|j)=1$ for both outcomes $j=1,2$. Then
\begin{align}
\nonumber
p_s&=\frac{1}{3}\sum_{l=1}^3 [q(l|1) x_l+ q(l|2)(1-x_l)]\\
\nonumber
&= \frac{1}{3}[1+\sum_{l=1}^3 [q(l|1)-q(l|2)]x_l]\\
&\leq \frac{1}{3}[1+\max_l x_l - \min_l x_l]<\frac{2}{3}=p_s^{\rm opt}\,,
\end{align}
because $0\leq\min_l x_l\leq \sum_l q(l|k) x_l\leq \max_l x_l\leq 1$ and
for the considered operators $M_{lk}$ we have strict inequality
$(\max_l x_l - \min_l x_l)<1$. In fact, the maximum probability of success $p_s=(2+\sqrt{3})/6$ for simple scheme
strategies is achieved for pure state $\varrho=\ket{\xi}\bra{\xi}$, where $\ket{\xi}=\cos{\omega}\ket{0}+\sin{\omega}\ket{1}$ and $\omega\approx 0.0833 \pi$.

\section{Unambiguous discrimination of two trine measurements}
\label{sec:trine}
Based on our previous analysis it is natural to ask whether the ancilla-based test procedures with maximally entangled states are always the ones (although not the unique ones) optimizing the discrimination figures of merits. In this section we will demonstrate an example rejecting such hypothesis.

\begin{figure}[t]
    \includegraphics[width=4.5cm ]{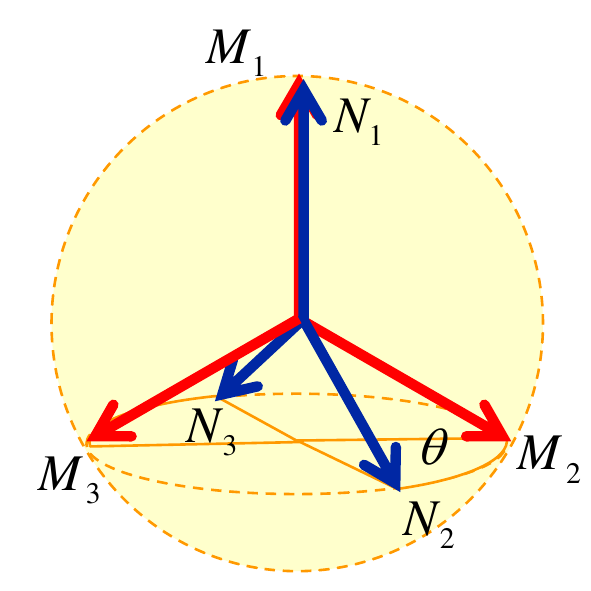}
    \caption{    \label{fig:mbpic} Two symmetric 3-outcome qubit POVMs in the Bloch representation that are mutually rotated by angle $\theta$, with respect to axis $z$.}
\end{figure}

Consider a symmetric three-outcomes qubit measurement
\begin{align}
\label{eq:twotrines}
\mM:\ M_1&=\frac{2}{3} \ket{0}\bra{0} \,, M_2=\frac{2}{3} \ket{v_+}\bra{v_+} \,, M_3=\frac{2}{3} \ket{v_-}\bra{v_-}\,,
\end{align}
where $\ket{v_\pm}=\frac{1}{2}\ket{0} \pm \frac{\sqrt{3}}{2}\ket{1}$.
Rotating this measurement by an angle $\theta$ around the $z$ axis
(see Fig. \ref{fig:mbpic}) we obtain a measurement $\mN_\theta$
with POVM elements $N_j= R_\theta M_j R^\dagger_\theta$, where
$R_\theta=\ket{0}\bra{0}+ e^{i\theta}\ket{1}\bra{1}$. In what follows we will
show that maximally entangled states as test states do not optimize success
probability for unambiguous discrimination of measurements
$\mM$ and $\mN_\theta$.

In the following we will use the lower bound
on the failure probability of unambiguous
discrimination of two channels $\mM$ and $\mN$ (with Choi operators
$M$ and $N$, respectively)
\begin{align}
p_f \ge 2\sqrt{\eta_\mM\eta_\mN}
{\rm tr} |\sqrt{M}(I\otimes \rho)\sqrt{N}|,
\label{eq:boundz1}
\end{align}
where $\rho\geq 0,\tr{\rho} =1$ is a normalization of the test used for the discrimination (see Eq. (\ref{def:norm1})).
The above bound was derived by Ziman et.al. in \cite{ziman1} (see Eq.(16) therein).
Our aim is to evaluate the bound for any normalization $\rho$ and to show that the bound can be saturated.
This will allow us to compare attainable failure probability for unambiguous discrimination of measurements
$\mM$ and $\mN_\theta$ for ancilla-based tests with maximally entangled states and those with optimal bipartite input states.

For uniform prior probabilities $\eta_\mM=\eta_{\mN_\theta}=1/2$ the bound (\ref{eq:boundz1}) reads
\begin{align}
\label{eq:lbtrine0}
p_f \ge \frac{3}{2}
{\rm tr} |M(I\otimes \rho)N_\theta|\,,
\end{align}
where we used the fact that $\sqrt{M}=\sqrt{3/2}\, M$,
$\sqrt{N_\theta}=\sqrt{3/2}\, N_\theta$ are the Choi operators of the
measurements $\mM$ and $\mN_\theta$, respectively.

Combining the triangle inequality for trace norm and the invariance of the norm
with respect to $\sigma_z$ rotation for the term with $j=3$ we can write the inequality
\begin{align}
\label{eq:lbtrines1}
\gamma=& {\rm tr}|M(I\otimes\rho)N_\theta|=\sum_j {\rm tr}|M^T_j \rho N^T_j| \\
\geq& {\rm tr}|M^T_1 \rho N^T_1| + {\rm tr}| M^T_2 \rho N^T_2 + \sigma_z M^T_3 \rho N^T_3 \sigma_z |\,. \nonumber
\end{align}
Using the parametrization
$\rho=q\ket{0}\bra{0}+(1-q)\ket{1}\bra{1}+z\ket{0}\bra{1}+z^*\ket{1}\bra{0}$
the above inequality reads
\begin{align}
\label{eq:lbcalc1}
\gamma
\geq \frac{4}{9}q+\frac{2}{9}\sqrt{q^2+ 9(1-q)^2+6p(1-q) \cos \theta}    \,.
\end{align}
Interestingly, this expression does not depend on $z$, hence, the only relevant parameter of $\rho$
is $q$. Combining Eqs. (\ref{eq:lbtrine0}), (\ref{eq:lbcalc1}) we get
\begin{align}
\label{eq:lbtrinep}
p_f \geq  \frac{2q + \sqrt{q^2+9(1-q)^2+6q(1-q) \cos \theta} }{3}\, ,
\end{align}
where $0 \leq q\leq 1$.

\begin{figure}[t]
    \includegraphics[width=8cm ]{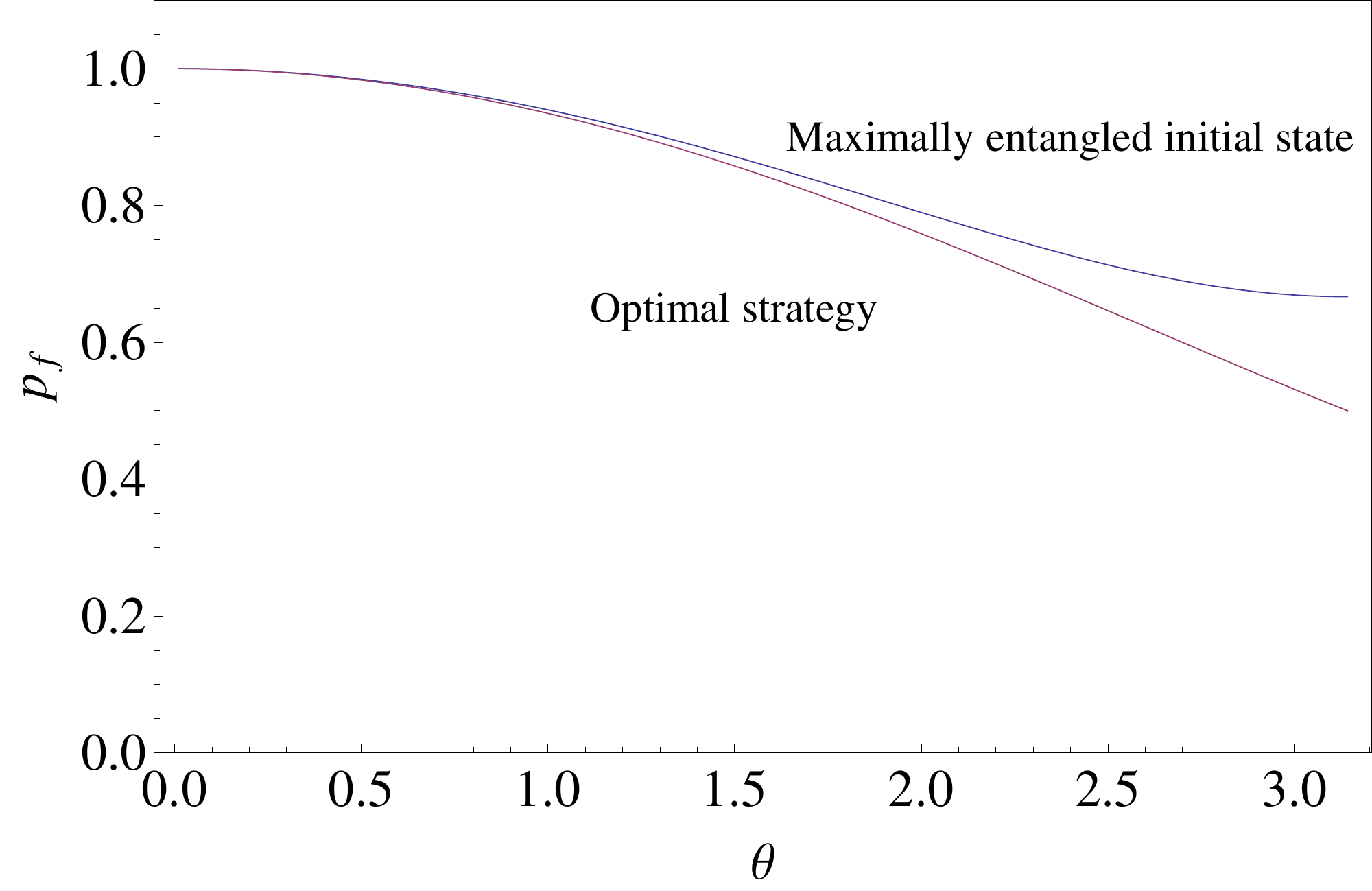}
    \caption{    \label{fig:graph1} Illustration of the difference between maximally entangled and optimal input bipartite state for the discrimination of two symmetric 3-outcome qubit POVMs mutually rotated by angle $\theta$, with respect to axis $Z$.}
\end{figure}

Next we consider a test procedure with normalization $\rho=q\ket{0}\bra{0}+(1-q)\ket{1}\bra{1}$, which saturates the above lower bound for every $0\leq q \leq 1$.
Consider a test state
\begin{align}
\ket{\phi_q}=\sqrt{q}\ket{00}+ \sqrt{1-q}\ket{11}.
\end{align}
Performing a trine measurement (either $\mM$, or $\mN_\theta$) on one of the qubits the second one ends up either in a conditional state $\ket{\psi_j^\mM}$, or $\ket{\psi_j^{\mN}}$. For $j=1$ these conditional states coincide with $\ket{0}$, thus, this outcome is necessarily inconclusive. The pairs of states to be discriminated for outcome $2$ and $3$ are mutually related by unitary transformation $\sigma_z$, so they have the same overlap
\begin{align}
\label{eq:trineoverlap1}
F=|\bra{\psi_2^\mM}\psi_2^\mN\>|=|\bra{\psi_3^\mM}\psi_3^\mN\>|=\frac{\big{|}q+e^{i\theta}(1-q)\big{|}}{3-2q}\,.
\end{align}
Using the results of Ivanovic \cite{idp1}, Dieks \cite{idp2} and Peres \cite{idp3} such pairs of pure equiprobable states can be unambiguously discriminated with failure probability equal to their overlap $F$. Weighting these cases by $p_j=\bra{\phi_q}M_j\ket{\phi_q}=\bra{\phi_q}N_j\ket{\phi_q}$, the probability of appearance of outcome $j$, we derive the average failure probability of the scheme
\begin{align}
\label{eq:lbtrines2}
p_f = \frac{2}{3}q + 2 \frac{3-2q}{6}F .
\end{align}
Inserting Eq.(\ref{eq:trineoverlap1}) into (\ref{eq:lbtrines2}) we see that the proposed scheme saturates the lower bound on the failure probability (\ref{eq:lbtrinep}) for any $q\in [0,1]$.
Thus, tuning $q$ in order to minimize the failure probability of the proposed scheme simultaneously gives the lowest achievable failure probability in general.
It can be shown that the minimum of the right hand side of Eq. (\ref{eq:lbtrinep}) is achieved for $q=(9-2\sqrt{3}\cos{(\theta/2)}-3\cos{\theta})/(10-6\cos{\theta})$ implying that
\begin{align}
\label{eq:lbtrinefinal}
p_f=\frac{1}{3} \Big(  1+\sqrt{3}\, \Big{|}\cos{\frac{\theta}{2}}\Big{|} + \frac{4-2\sqrt{3}\;|\cos{\frac{\theta}{2}}|}{5-3\cos{\theta}} \Big{)}.
\end{align}
Finally, let us assume that the test state is any maximally entangled state.
Any such test has normalization $\rho=\frac{1}{2}I$ corresponding to $q=1/2$.
Thus, by comparing the failure probability given by Eq. (\ref{eq:lbtrinep}) for $q=1/2$ and for $q$ minimizing the failure probability we can demonstrate that the use of less than maximally entangled states is needed in order to achieve the optimal performance. The difference is illustrated in Figure \ref{fig:graph1}.

\section{Summary}
\label{sec:summary}
In this paper we studied discrimination of quantum measurements with finitely many outcomes in the scenario when the unknown measurement can be used only once, but use of any other resources is allowed. In particular, we investigated special instances of the discrimination with fixed failure rate. This class of problems includes perfect discrimination, minimum-error discrimination and unambiguous discrimination.

We studied first the conditions for perfect discrimination. We have shown that the maximal number of distinguishable measurement apparatuses is bounded by the total number of outcomes $n$.
Let us stress that the dimension of the system is irrelevant and
one can find arbitrarily many qubit observables that are single-shot perfectly distinguishable.
Further, we have formulated a reduction theorem excluding a subspace irrelevant for the discrimination. More precisely, we showed that any subspace common to a given outcome of both measurements is irrelevant for the discrimination. We employed this theorem to relate the discrimination of quantum filters to discrimination of projective qubit measurements.

We found that the optimization of the discrimination of projective qubit measurements is mathematically equivalent to solving discrimination of pure states. Not only the optimal success rates are the same, but also the optimal discrimination algorithm for pure state discrimination can be directly exploited for optimal discrimination of projective qubit measurements. First, we prepare a singlet state of two qubits and apply the unknown measurement on one part of the state. This projects the second qubit into a pure state determined by the obtained outcome and the identity of the unknown measurement. Conditionally on the observed outcome we employ the optimal state discrimination strategy to identify the projected state of the second qubit, hence,
identifying the measurement used. Using this "measurement-to-state" reduction we provide solution to optimal minimum-error and unambiguous discrimination of projective qubit measurements. Let us note that this
procedure was successfully experimentally implemented in quantum optical system \cite{dmolomouc}. We extent this result to  the case of $m$ projective qubit measurements and, in addition, each of them may be affected by different level of white noise. Unfortunately, we have not succeeded to formulate similar result in more dimensional Hilbert spaces, where already the optimal discrimination of projective measurements is left open.

Our results clearly exhibits the added value of maximally entangled states although we have argued that in the case of perfect discrimination of binary measurements the ancilla can be completely ignored and simple scheme works as well as the entangled one. From the algebraic point of view the measurements are channels mapping quantum (non-commutative) algebra to classical (commutative) one, hence, the concepts of positivity and complete positivity coincide, i.e. tensor product extensions of such channels are irrelevant for judging this property. However, we were surprised to find an example exhibiting the fact that even in case of (perfect) discrimination between only a pair of measurements, the ancilla, hence, tensor product extensions of the channels, provides an advantage over the simple (ancilla-free) schemes. It is an intriguing question to understand in which cases the simple scheme performs as good as the general one, and when the maximally entangled states provide the optimal discrimination strategy. We have shown an explicit example demonstrating situations in which non-maximally entangled states 
outperform maximally entangled ones.

\acknowledgments
M.S. acknowledges support by the Operational Program Education for Competitiveness - European Social Fund (project No. CZ.1.07/2.3.00/30.0004) of the Ministry of Education, Youth and Sports of the Czech Republic.  M.Z. acknowledges the support of projects VEGA 2/0125/13 (QUICOST), APVV-0646-10 (COQI) and
GA\v CR P202/12/1142.

\appendix

\section{Proof of lemma \ref{lem:reduction1}}
\label{sec:appreduction}
Suppose that a test procedure $\mathcal{T}$ specified by operators $H^{(c)}_j$ $ c \in \{\mathcal{M},\mathcal{N},f\}$, $j\in \Omega$ leads to certain values of $p_s,p_e$ and $p_f$.
In the first step our aim is to design a different test procedure $\mathcal{T'}$ that would give the same values of $p_s,p_e$ and $p_f$
that can be interpreted as a mixture of the subproblem defined in the section \ref{sec:redtheorem} and a discrimination of two identical measurements.

We define
$H'^{(c)}_j=\sum_{k\in \omega} Q_k H^{(c)}_j Q_k + (1-P)H^{(c)}_j(1-P)$,
where we recall $P=\sum_{k\in \omega} Q_k$. By definition operators $H'^{(c)}_j$ are positive semidefinite and they obey the following normalization
\begin{align}
\forall j \quad \sum_c H'^{(c)}_j
=\sum_k Q_k\rho Q_k+(1-P)\rho(1-P) \equiv \rho', \nonumber
\end{align}
where we defined positive semidefinite operator $\rho'$.
Moreover, $\tr{\rho'}=1$, so we showed that operators $H'^{(c)}_j$ specify a valid test procedure.

Due to $M_k \geq Q_k$ and $\sum_k M_k=I$ we have
$M_k=I-\sum_{l\neq k} M_l\leq I- \sum_{l\neq k} Q_l$, which is equivalent to
\begin{align}
\label{eq:appopeq1}
M_k-Q_k\leq I-P.
\end{align}
As a consequence,
\begin{align}
\label{eq:appopeq2}
Q_l M_k Q_l=\delta_{kl} Q_l,
\end{align}
because for $k\neq l$ we get $0\leq Q_l M_k Q_l\leq 0$ and case $k=l$ follows from the definition of $Q_k$.
Finally, using Eq. (\ref{eq:appopeq1}) we get
$(1-P)(M_k-Q_k)(1-P)=M_k-Q_k$, which is useful to write as:
\begin{align}
\label{eq:redlema2}
M_k=Q_k M_k Q_k+(1-P)M_k(1-P),
\end{align}
where we used the above identities and $Q_k Q_l=\delta_{kl} Q_l$.

Analogously one can derive
relations (\ref{eq:appopeq2}),(\ref{eq:redlema2}) for elements $N_i$.
This enables us to show that the test procedure $\mathcal{T'}$ leads to the same values of $p_s,p_e$ and $p_f$, because the conditional probabilities
$p(c|\mathcal{M},\mathcal{T})$, $p(c|\mathcal{N},\mathcal{T})$ do not change.
Indeed, we have
\begin{align}
\label{eq:redlema1}
p(c|\mathcal{M},\mathcal{T'})
&= \sum_{j}\sum_k \tr{Q_k H^{(c)}_j Q_k M_j} \\
&\quad + \sum_{j} \tr{(1-P)H^{(c)}_j (1-P) M_j} \nonumber \\
&=\sum_{j} \tr{ H^{(c)}_j Q_j M_j Q_j} \nonumber \\
&\quad + \sum_{j} \tr{ H^{(c)}_j (1-P) M_j(1-P)} \nonumber \\
&=\sum_j \tr{H^{(c)}_j M_j}=p(c|\mathcal{M},\mathcal{T}), \nonumber
\end{align}
where we used Eqs. (\ref{eq:appopeq2}),(\ref{eq:redlema2}).
Analogously one can show $p(c|\mathcal{N},\mathcal{T'})=p(c|\mathcal{N},\mathcal{T})$.
Let us introduce Hilbert spaces $\widetilde{\hilb{H}}=(1-P)\hilb{H}$, $\overline{\hilb{H}}=P\hilb{H}$ specified by the projector $P$ and its complement.
From the assumptions of the theorem we have that $\widetilde{\mM}$ and $\widetilde{\mN}$ form a measurement on the Hilbert space $\widetilde{\hilb{H}}$.
If $P\rho P =0$ then $H'^{(c)}_j=H^{(c)}_j$ and it already specifies a discrimination procedure for $\widetilde{\mM}$ and $\widetilde{\mN}$ in $\widetilde{\hilb{H}}$.
Similarly, if $(1-P)\rho (1-P)=0$ then $H'^{(c)}_j$ specifies a discrimination procedure for
$\overline{\mM}=\overline{\mN} \leftrightarrow \{Q_j\}_{j=1}^n$
in $\overline{\hilb{H}}$.
In the rest of the cases we define $\lambda=\tr{P\rho}$ and
\begin{align}
\label{eq:defsubtests}
\widetilde{\rho}&=\frac{(1-P)\rho (1-P)}{1-\lambda} \quad  &\overline{\rho}&=\frac{1}{\lambda}\sum_k Q_k\,\rho\, Q_k  \\
\widetilde{H}^{(c)}_j&=\frac{(1-P)H^{(c)}_j(1-P)}{1-\lambda} \quad &\overline{H}^{(c)}_j&=\frac{1}{\lambda}\sum_k Q_k\, H^{(c)}_j Q_k  \nonumber
\end{align}
It is now easy to see that operators $\widetilde{H}^{(c)}_j$ describe a valid discrimination procedure $\widetilde{\mathcal{T}}$ for $\widetilde{\mM}$ and $\widetilde{\mN}$, while
$\overline{H}^{(c)}_j$ do the same for $\overline{\mM}$ and $\overline{\mN}$.
Using definitions (\ref{eq:defsubtests}) the conditional probabilities can be rewritten  (see also Eq. (\ref{eq:redlema1})) as
\begin{align}
p(c|\mathcal{M},\mathcal{T'})&=(1-\lambda) \sum_{j} \tr{\widetilde{H}^{(c)}_j M_j} + \lambda \sum_{j} \tr{\overline{H}^{(c)}_j M_j} \nonumber \\
&=(1-\lambda ) p(c|\widetilde{\mathcal{M}},\widetilde{\mathcal{T}}) + \lambda p(c|\overline{\mathcal{M}},\overline{\mathcal{T}})
\end{align}
As a consequence, we have
\begin{align}
p_s &=(1-\lambda) \widetilde{p}_s + \lambda \overline{p}_s \nonumber \\
p_e &=(1-\lambda) \widetilde{p}_e + \lambda \overline{p}_e  \\
p_f &=(1-\lambda) \widetilde{p}_f + \lambda \overline{p}_f. \nonumber
\end{align}
Thus, performance of any test $\mathcal{T}$ can be also attained by a suitable test $\mathcal{T}'$, which naturally defines operators $\widetilde{H}^{(c)}_j$, $\overline{H}^{(c)}_j$ for
discrimination of measurements $\widetilde{\mM}$, $\widetilde{\mN}$ and $\overline{\mM}$, $\overline{\mN}$, respectively.
Moreover, also the opposite holds, i.e. every properly normalized set of operators $\widetilde{H}^{(c)}_j$, $\overline{H}^{(c)}_j$ and a coefficient $0\leq \lambda \leq 1$ defines a valid test $\mathcal{T}'$.

Next, we want to show that in order to maximize probability of success $p_s$ it suffices to consider tests $\mathcal{T}''$ with $\lambda=0$ i.e. $P\rho''P =0$.
Such tests of the unknown measurement use input states that do not probe the subspace of the Hilbert space $\hilb{H}$ defined by projector $P$.

Since any pair of measurements can be discriminated at least as good as two indistinguishable measurements
we can find operators $\widehat{H}^{(c)}_j$ for discrimination of  $\widetilde{\mM}$, $\widetilde{\mN}$ with failure probability $\widehat{p}_f=\overline{p_f}$ and $\widehat{p}_s\geq \overline{p}_s$.
A test $\mathcal{T}''$ defined by operators $H''^{(c)}_j=(1-\lambda)\widetilde{H}^{(c)}_j+ \lambda\widehat{H}^{(c)}_j$ has the same failure probability $p_f$ as the test $\mathcal{T}'$, but it has a higher success probability $(1-\lambda)\widetilde{p}_s+\lambda\widehat{p}_s \geq p_s$. Moreover, $P H''^{(c)}_j P=0$ implies $P\rho''P =0$, so we showed that it suffices to consider only tests $\mathcal{T}''$ or in other words it suffice to solve the discrimination problem for measurements $\widetilde{\mM}$, $\widetilde{\mN}$ instead of the original problem. The optimal discrimination procedure is the same in both cases except for being formally defined on a bigger Hilbert space.

\end{document}